\documentclass[11pt]{article}
\usepackage{jheppub}

\usepackage{color}
\usepackage{amsmath}
\usepackage{verbatim}   
\usepackage{subfigure}  
\usepackage{acronym}

\usepackage{amsfonts}
\usepackage{amssymb}
\usepackage{mathrsfs}
\usepackage{graphicx}
\usepackage{multirow}
 \usepackage{slashed}
 \usepackage{epsfig,multicol,bbm}
 \usepackage{url}

\newcommand{\newc}{\newcommand}
\newc{\gsim}{\lower.7ex\hbox{$\;\stackrel{\textstyle>}{\sim}\;$}}
\newc{\lsim}{\lower.7ex\hbox{$\;\stackrel{\textstyle<}{\sim}\;$}}
\newc{\gev}{\,{\rm GeV}}
\newc{\mev}{\,{\rm MeV}}
\newc{\ev}{\,{\rm eV}}
\newc{\kev}{\,{\rm keV}}
\newc{\tev}{\,{\rm TeV}}

\newc{\mz}{M_Z}
\newc{\mpl}{M_*}
\newc{\mw}{m_{\rm weak}}
\newc{\nr}[1]{N^c_R{}_{#1}}
\usepackage{amsmath}

\def\beq{\begin{equation}}
\def\eeq{\end{equation}}
\def\bea{\begin{eqnarray}}
\def\eea{\end{eqnarray}}
\def\bitem{\begin{itemize}}
\def\eitem{\end{itemize}}
\newcommand{\bec}{\begin{center}}
\newcommand{\eec}{\end{center}}
  \newcommand{\GeV}{{\mathrm {GeV}}}

     \newcommand{\fb}{{\mathrm {fb}}}
        
\def\bar#1{\overline{#1}}

\def\inv{^{\raise.15ex\hbox{${\scriptscriptstyle -}$}\kern-.05em 1}}
\def\lbar{{\lower.35ex\hbox{$\mathchar'26$}\mkern-10mu\lambda}} 

\let\<=\langle
\let\>=\rangle

\let\+=\uparrow

\let\la=\lambda
\let\La=\Lambda

\def\OO{\mathcal{O}}

\begin{document}

\hfill \vspace{-5mm} OUTP-12-01P

\title{Closing in on Asymmetric Dark Matter I:\\
Model independent limits for interactions with quarks}

\author[a]{John March-Russell,}
\emailAdd{jmr@thphys.ox.ac.uk}

\author[a,b]{James Unwin}
\emailAdd{unwin@maths.ox.ac.uk}

\author[c,d]{and Stephen M. West}
\emailAdd{stephen.west@rhul.ac.uk}

\affiliation[a]{Rudolf Peierls Centre for Theoretical Physics,
University of Oxford,\\
1 Keble Road, Oxford,
OX1 3NP, UK}

\affiliation[b]{Mathematical Institute,
University of Oxford,
24-29 St Giles, Oxford,
OX1 3LB, UK}

\affiliation[c]{Royal Holloway, University of London, Egham, TW20 0EX, UK}

\affiliation[d]{Rutherford Appleton Laboratory, Chilton, Didcot, OX11 0QX, UK}

\date{\today}

\abstract{It is argued that experimental constraints on theories of asymmetric dark matter (ADM) almost certainly require that the DM be part of a richer hidden sector of interacting states of comparable mass or lighter. 
A general requisite of models of ADM is that the vast majority of the symmetric component of the DM number density must be removed in order to explain the observed relationship $\Omega_B\approx5\Omega_{DM}$ via the DM asymmetry. Demanding the efficient annihilation of the symmetric component leads to a tension with experimental limits if the annihilation is directly to Standard Model (SM) degrees of freedom. 
A comprehensive effective operator analysis of the model independent constraints on ADM from direct detection experiments and LHC monojet searches is presented. 
Notably, the limits obtained essentially exclude models of ADM with mass $1~{\rm GeV}\lesssim m_{\mathrm{DM}} \lesssim 100~{\rm GeV}$ annihilating to SM quarks via heavy mediator states. This motivates the study of portal interactions between the dark and SM sectors mediated by light states. Resonances and threshold effects involving the new light states are shown to be important for determining the exclusion limits.}

\maketitle


\section{Introduction}

Asymmetric dark matter (ADM) provides a compelling framework for explaining the existence and comparable magnitudes of the present day baryon density $\Omega_B$ and the dark matter (DM) relic density $\Omega_{\mathrm{DM}}$. In traditional cosmology the origins of $\Omega_B$ and $\Omega_{\mathrm{DM}}$ are distinct and unrelated. Baryogenesis is conventionally attributed to CP violating processes in baryons or leptons, whilst $\Omega_{\mathrm{DM}}$ is assumed to be determined by particle species decoupling from the thermal bath (freeze-out) \cite{FO, FO2, FO3}, or out-of-equilibrium species moving towards thermal equilibrium with the visible sector (freeze-in) \cite{Hall:2009bx}. Thus, a priori, the DM and baryon densities could have differed by orders of magnitude and the notably close proximity  $\Omega_{\mathrm{DM}}\approx4.56\times\Omega_B$ has motivated a concentrated research  effort to connect the genesis mechanisms for baryons and DM \cite{Nussinov:1985xr, Gelmini, Chivukula:1989qb, Barr:1990ca, Kaplan:1991ah, Hooper:2004dc, Kitano:2004sv, Cosme:2005sb, Farrar:2005zd, Suematsu:2005kp, Tytgat:2006wy, Kaplan:2009ag}.

For the DM to be stable and linked to the baryon asymmetry, in addition to the observed quantum number which stabilises the state (e.g. $\mathbb{Z}_2$ $R$-parity), the DM must carry an additional conserved quantum number analogous to baryon number, which we denote $X$. At low energies $\mathrm{U}(1)_B$ and $\mathrm{U}(1)_X$ are approximate global symmetries of the visible and hidden sector, respectively. However, at higher energies these are no longer good symmetries and the only conserved quantity is $B-L+X$. Consequently, models can be constructed in which asymmetries in baryon, lepton and DM number are linked, and thus provide a natural explanation for the relative coincidence of the cosmic relic densities. There now exist a large variety of mechanisms for connecting DM with baryogensis \cite{An:2009vq, Arina, Kaplan:2009de, Shelton:2010ta, Haba:2010bm, Buckley:2010ui, Chun:2010hz, Gu:2010ft, Dutta:2010va, Blennow:2010qp, Kitano:2008tk, Falkowski:2011xh, D'Eramo:2011ec, Kang:2011ny, Kohri, Belyaev:2010kp, Frandsen:2011kt, Cui:2011qe, Hall:2010jx, Davoudiasl:2010am, Cheung:2011if,MarchRussell:2011fi}, but they roughly fall into two broad categories: sharing and cogenesis.  In models of sharing one begins with a pre-existing asymmetry in some (approximately) conserved quantum number and postulates a connector operator through which the asymmetry may be transferred among $B-L$ and $X$. In the cogenesis scenario one assumes that there is no initial asymmetry and that the asymmetries in baryons and DM are simultaneously generated via some out-of-equilibrium CP violating process  \cite{Kitano:2008tk,Hall:2010jx, Davoudiasl:2010am, Cheung:2011if,MarchRussell:2011fi}.  In many models of cogenesis the baryon and DM asymmetries are equal and opposite, such that there is no overall asymmetry in the quantity $B-L+X$. Since natural realisations of ADM lead to comparable asymmetries in $B$ and $X$, to account for $\Omega_B/\Omega_{\mathrm{DM}}\approx5$, the DM mass $m_{\mathrm{DM}}$ must be of similar magnitude to the proton mass $m_p\approx  0.94$ GeV. We shall refer to the range $1~{\rm GeV}\lesssim m_{\mathrm{DM}} \lesssim 10~{\rm GeV}$ as the ADM mass range and focus much of our attention on this mass region. Although the ADM mass range is preferred theoretically, we include limits on model of ADM with DM mass in the range to $1~{\rm GeV}\lesssim m_{\mathrm{DM}} \lesssim 10^4~{\rm GeV}$.

In standard theories of baryogenesis an asymmetry $\eta_B$ is generated between the baryons and anti-baryons which ultimately leads to the visible universe. Due to the strong nuclear force the symmetric component of the baryon density pair-wise annihilates, leaving only the asymmetric component of the baryons, resulting in the matter dominated universe we observe.  The framework of ADM proposes that the (non-self-conjugate) DM undergoes a similar cosmological history to the baryons and the DM relic density is due to an asymmetry between the DM and its antiparticle. Similar to baryons, the DM must have sufficiently large interactions such that only a negligible symmetric component remains, in which case the baryon and DM relic densities are related by
\begin{equation}
\Omega_{\mathrm{DM}}\simeq \frac{m_{\mathrm{DM}}\eta_{\mathrm{DM}}}{m_p\eta_B}\Omega_{B}.
\end{equation}
To link $\Omega_B$ and $\Omega_{\mathrm{DM}}$ in this manner it is essential that the DM relic density be due to the DM  {\em{asymmetry}} and consequently, as discussed in \cite{Hall:2010jx}, it is a generic requirement that the vast majority of the symmetric component is removed. 

In this paper we shall remain agnostic as to the source of the DM asymmetry, instead focusing on a general feature of ADM from which model independent bounds can be derived. Whilst DM annihilation is more efficient for larger values of the inter-sector couplings, it is precisely these inter-sector couplings which are constrained through direct detection and collider searches.  Hence, the need for efficient annihilation results in a tension between cosmological requirements and direct search constraints.  Generically, the symmetric part of the DM will be several orders of magnitude larger than the asymmetric component, as the ratio of asymmetric to symmetric DM is of comparable magnitude to the CP violating parameter determining the asymmetry and is typically $\lesssim10^{-3}$. As a result, these constraints are sufficiently strong to rule out large regions of parameter space. 


\begin{table}[t!]
\begin{center}
\begin{tabular}{|l|c|c|}
\hline
$ \hspace{1.5cm}\Delta\mathscr L$& Int. & Suppression
\\
\hline
$ \OO^\phi_{s}:\quad\frac{1}{\La}\phi^{\dagger} \phi \bar f f $ &SI& \(1\)\\[3pt]
$  \OO^\phi_{v}:\quad\frac{1}{\La^2} \phi^{\dagger} \partial^\mu \phi \bar f \gamma_\mu f $&SI& \(1\)\\[3pt]
$  \OO^\phi_{va}:\quad\frac{1}{\La^2} \phi^{\dagger} \partial^\mu \phi \bar f \gamma_\mu\gamma^5 f $ &SD& \(v^2\)\\[3pt]
$  \OO^\phi_{p}:\quad\frac{1}{\La} \phi^{\dagger} \phi \bar f i\gamma^5 f $&SD& \(q^2\)\\[3pt]
\hline
$\OO^\psi_{s}:\quad\frac{1}{\La^2} \bar \psi \psi \bar f f $ &SI& \(1\)\\[3pt]
$\OO^\psi_{v}:\quad\frac{1}{\La^2} \bar \psi \gamma^{\mu} \psi \bar f \gamma_{\mu} f $ &SI&\(1\)\\[3pt]
$ \OO^\psi_{a}:\quad\frac{1}{\La^2}\bar \psi \gamma^{\mu}\gamma^5 \psi \bar f \gamma_{\mu} \gamma^5 f $ &SD&\(1\)\\[3pt]
$ \OO^\psi_{t}:\quad\frac{1}{\La^2}\bar \psi \sigma^{\mu \nu} \psi \bar f \sigma_{\mu \nu} f$&SD&\(1\)\\[3pt]
$ \OO^\psi_{p}:\quad\frac{1}{\La^2}\bar \psi \gamma^5 \psi \bar f \gamma^5 f $ &SD& \(q^4\)\\[3pt]
$\OO^\psi_{va}:\quad\frac{1}{\La^2} \bar \psi \gamma^{\mu} \psi \bar f \gamma_{\mu} \gamma^5 f $ &SD&\(v^2\), \(q^2\)\\[3pt]
$\OO^\psi_{pt}:\quad\frac{1}{\La^2}\bar \psi i\sigma^{\mu\nu} \gamma^5 \psi \bar f \sigma_{\mu\nu} f $&SI& \(q^2\)\\[3pt]
$\OO^\psi_{ps}:\quad\frac{1}{\La^2} \bar \psi i\gamma^5 \psi \bar f f $&SI& \(q^2\)\\
$\OO^\psi_{sp}:\quad\frac{1}{\La^2}\bar \psi \psi \bar f i\gamma^5 f $&SD& \(q^2\)\\
\hline
\multirow{2}{*}{$\OO^\psi_{av}:\quad\frac{1}{\La^2}\bar \psi \gamma^{\mu} \gamma^5\psi \bar f \gamma_{\mu}  f $}&SI& \(v^2\)\\
                                                                                                                                  & SD &  \(q^2\) \\
\hline
$ \hat\OO^\phi_{s}:\quad\frac{m_q}{\La^2}\phi^{\dagger} \phi \bar f f $ &SI& \(1\)\\[3pt]
$\hat\OO^\psi_{s}:\quad\frac{m_q}{\La^3} \bar \psi \psi \bar f f $ &SI& \(1\)\\[3pt]
$ \hat\OO^\psi_{p}:\quad\frac{m_q}{\La^3}\bar \psi \gamma^5 \psi \bar f \gamma^5 f $ &SD& \(q^4\)\\[3pt]
\hline
\end{tabular}
\end{center}
\caption{Set of effective operators $\OO$  of dimension 6 or less connecting SM quarks $f$ to scalar $\phi$ or fermion $\psi$ DM with universal couplings. We consider 3 additional natural contact operators $\hat\OO$ with quark mass dependent couplings, note $\hat\OO\equiv\frac{m_q}{\La}\OO$. It is indicated whether the corresponding direct detection cross-section is DM velocity $v$ or momentum transfer $q$ suppressed and if the interaction couples to nuclei in a spin-dependent (SD) or independent (SI) manner \cite{Freytsis:2010ne}.}
\label{tab1}
\end{table}


Since momentum transfer is low in  direct detection experiments, mediators with masses $\sim100$ MeV or greater are sufficiently heavy to be integrated out. Consequently the variety of portal interactions between the hidden and visible sectors may be parameterised via effective contact operators.   We shall demonstrate that the current experimental limits exclude ADM models in which the DM couples to SM quarks with minimal flavour structure via a heavy mediator, for $1~{\rm GeV}\lesssim m_{\mathrm{DM}} \lesssim 100~{\rm GeV}$ . However, effective operators do not provide a good description of the collider limits and the relic density calculation for mediators with masses $\lesssim100$ GeV and, hence, in Section \ref{light} we examine the effects of such light mediators. In Table \ref{tab1} we list all effective operators up to dimension 6 connecting SM quarks $f$ to complex scalar $\phi$ or Dirac fermion $\psi$ DM.\footnote{Since the DM must carry a conserved U$(1)_{X}$ Majorana fermions or real scalars are not suitable candidates for ADM. Recall  $\sigma^{\mu\nu}=\frac{i}{2}(\gamma^\mu\gamma^\nu-\gamma^\nu\gamma^\mu)$ and since $ \sigma^{\mu\nu}\gamma^5 =  \frac{i}{2} \epsilon^{\mu\nu\rho\sigma}\sigma_{\rho\sigma} $, the operators $\bar \psi i\sigma^{\mu\nu} \gamma^5 \psi \bar f \sigma_{\mu\nu}f$ and $ \bar \psi \sigma^{\mu \nu} \psi \bar f \sigma_{\mu \nu}f$ are equivalent to $\bar \psi i\sigma^{\mu\nu}\psi \bar f \sigma_{\mu\nu} \gamma^5  f $ and $ \bar \psi \sigma^{\mu \nu}  \gamma^5 \psi \bar f \sigma_{\mu \nu}  \gamma^5 f$, respectively.}  
 For definitiveness and to avoid stringent flavour-changing constraints we will typically assume that the contact operators couple universally to all quark flavours. In the case of scalar and pseudoscalar interactions we shall also study the case where the effective coupling to SM quarks is proportional to the relevant quark mass, which is arguably more natural, and we denote these $m_q$-dependent operators $\hat\OO\equiv\frac{m_q}{\La}\OO$.  Operators $ \OO^\phi_{p}$, $ \OO^\psi_{pt}$, $ \OO^\psi_{ps}$ and $ \OO^\psi_{sp}$ correspond to CP violating interactions and one might expect their coefficients to be suppressed relative to those of the other, CP conserving, operators, consequently we shall typically omit these contact interactions for brevity. We have checked that the requirements for efficient annihilation of the symmetric component and experimental limits for these CP violating operators are comparable to examples which will be presented in detail.

The direct detection cross-section of certain operators are suppressed by the small DM velocity  $v\sim 10^{-3}c$ or momentum transfer \(|q|=|p_i-p_f|\lesssim0.1\) GeV.  For operators with suppressed scattering cross-sections the current direct detection limits are insufficient to provide any useful constraints on the DM over the whole mass range. However, the high energy production cross-sections remain unsuppressed and, consequently, monojet searches provide the leading limits on such operators. 
Approaches based on effective operators have been employed previously to study the constraints from direct detection and colliders on conventional DM models \cite{Freytsis:2010ne, Kurylov:2003ra, Fitzpatrick:2010em, Fan:2010gt, Agrawal:2010fh, Beltran:2008xg, Fox:2011fx, Harnik:2008uu, Rajaraman:2011wf, Fox:2011pm, Friedland:2011za, Shoemaker:2011vi, Razor}, however, the existence of an asymmetry in the hidden sector leads to deviations in the relic density calculations \cite{Griest:1986yu}.  Whilst there have been previous papers on contact operators in the context of ADM \cite{Buckley:2011kk, Lin:2011gj}, in the case of the former the effects of the DM asymmetry are not fully accounted for, while the latter work considers only a limited selection of possible operators. In the present work we consider the full set of contact operators (correcting some numerical errors contained in previous works) and include the latest experimental constraints from both direct detection and LHC monojet searches (including the recent CMS results with 4.67$\fb^{-1}$). Moreover, in later sections, we go beyond the contact operator analysis and demonstrate that if the DM is connected to SM quarks via exchange of a light mediator state then resonance and threshold effects become very important in determining the exclusion limits.

As discussed previously, if the DM annihilates directly to SM quarks then there is a tension between the efficient removal of the symmetric component and experimental searches. However, if the DM is embedded into a richer hidden sector, with a spectrum of lighter states into which the symmetric component can annihilate (which later decay to the visible sector), then this severs the connection between direct search constraints and cosmological requirements. As we show that effective theories of ADM with $m_{\mathrm{DM}}\sim m_p$ and minimal flavour structure which decay directly to SM quarks are excluded for all types of connector operator, this leads to the striking conclusion that in order to realise ADM naturally, the DM must generally be accompanied by additional hidden states of comparable mass or lighter, either to mediate the interactions to the visible sector, or into which the symmetric component of the DM may annihilate. The fact that ADM must be part of a larger hidden sector is an important realisation, since this will generally lead to significant changes to the cosmology and phenomenology, which we shall address in a companion paper \cite{companion}.

This paper is structured as follows, in Section \ref{Sec2} we review the standard treatment for calculating the DM relic density in the presence of an asymmetry. In Section  \ref{Sec3}  we present a comprehensive study of effective operators and compare the requirements for successful ADM models to the experimental limits from direct detection and monojet searches. The results of Section \ref{Sec3} strongly motivate a careful study of light mediators and we present an analysis of the scalar and pseudoscalar portal interactions due to light mediators in Section \ref{light}. Finally, in the concluding remarks we discuss the implications these results have on model building with ADM and emphasise that the null experimental searches strongly imply that successful models must feature an extended hidden sector.


\section{Relic abundance with an asymmetry}
\label{Sec2}

The DM asymmetry alters the relic density calculation from the conventional case without an asymmetry.  Relic density calculations in the ADM paradigm were recently discussed in \cite{Iminniyaz:2011yp,Graesser:2011wi} and we shall follow these analyses in determining the relic density of the symmetric component. Generically, it was found in these analyses that the annihilation cross-section is required to be larger than in the case with no asymmetry.

We define the standard variable $x\equiv m/T$, for a particle of mass $m$ at temperature $T$ and  denote by $x_F$ the inverse scaled decoupling temperature of DM and antiDM, assuming that the only $X$ number changing interactions are pair-annihilations of the DM and its anti-partner. Following \cite{Iminniyaz:2011yp}, the yields for $X$ and $\bar{X}$ in the limit that $x\rightarrow\infty$ are given by:
\begin{equation}
\begin{aligned}
Y_{\mathrm{DM}} &= \frac{\eta_{\mathrm{DM}}}{1
-\exp\left[-\eta_{\mathrm{DM}} J \omega \right]},
\hspace{15mm}
Y_{\bar{\mathrm{DM}}} &= \frac{\eta_{\mathrm{DM}}}{\exp\left[ \eta_{\mathrm{DM}} J \omega\right]-1},
\label{YXJ}
\end{aligned}
\end{equation}
where $\eta_{\mathrm{DM}}\equiv Y_{\mathrm{DM}}-Y_{\bar{\mathrm{DM}}}$ is the asymmetric yield. The quantity $\omega$ is defined by
 \begin{equation}
\omega=\frac{4\pi}{\sqrt{90}}m_{\mathrm{DM}}M_{\mathrm{Pl}}\sqrt{g_{*}}
\end{equation}
where $m_{\mathrm{DM}}$ is the DM mass, $M_{\mathrm{Pl}}=2.4\times10^{18}$ is the reduced Plank mass and $g_*(T)$ is the number of available relativistic degrees of freedom at a given temperature $T$. 
The annihilation integrals $J$ are given by
\begin{equation}
J=\int_{x_F}^\infty\frac{\langle\sigma v \rangle}{x^2}\,\mathrm{d}x.
\end{equation}
Making an expansion of the annihilation cross-section\footnote{As discussed in Section \ref{light} and the Appendix, this expansion is not appropriate in the case of light mediators and must be replaced with an alternative analysis and numerical methods.}
\begin{equation}
\langle\sigma v \rangle= a+\frac{6b}{x}+\OO(x^{-2}),
\label{sig}
\end{equation}
allows us to express eqn. (\ref{YXJ}) as follows
\begin{equation}
\begin{aligned}
Y_{\mathrm{DM}} &= \frac{\eta_{\mathrm{DM}}}{1
-\exp\left[-\eta_{\mathrm{DM}} \omega\left(\frac{a}{x_F}+\frac{3b}{x_F^2}\right)\right]},
\hspace{10mm}
Y_{\bar{\mathrm{DM}}} &= \frac{\eta_{\mathrm{DM}}}{\exp\left[ \eta_{\mathrm{DM}}\omega\left(\frac{a}{x_F}+\frac{3b}{x_F^2}\right) \right]-1}.
\label{YX}
\end{aligned}
\end{equation}
Using the tables from DarkSUSY \cite{Gondolo:2004sc} for the temperature dependence of $g_*(T)$, we model the behaviour of $x_F$ via the approximation introduced in \cite{Iminniyaz:2011yp}:
\begin{equation}
 x_F=x_{F0}\left(1+0.285 \frac{a  \omega  \eta_{\mathrm{DM}}}{x_{F0}^3}
 +1.350\frac{b  \omega  \eta_{\mathrm{DM}}}{x_{F0}^4}\right),
\label{x}
 \end{equation}
where $x_{F0}$ is the inverse scaled decoupling temperature in standard freeze-out with no asymmetry.
The present relic density is given by
\begin{equation}
\Omega_{\mathrm{DM}}h^2=2.76\times 10^8 \left( \eta_{\mathrm{DM}}+Y_{\mathrm{Sym}} \right)\left(\frac{m_{\mathrm{DM}}}{\GeV}\right),
\end{equation}
where $Y_{\mathrm{Sym}}= 2Y_{\bar{\mathrm{DM}}}$ is the yield of the symmetric component and  $h=0.710\pm0.025$ is the Hubble constant in units of 100 $\mathrm{km}\mathrm{s}^{-1}\mathrm{Mpc}^{-1}$. In the limit $\eta_{\mathrm{DM}} \rightarrow0$, one recovers the standard symmetric DM expressions. The observed value of the DM abundance \cite{Jarosik:2010iu} is $\Omega_{\mathrm{DM}}h^2=0.1109\pm0.0056$. In order to realise the requirement that the asymmetric component constitutes the majority of the present relic density, we demand that the symmetric component  composes $\leq1\%$ of the relic density, via the constraint:
\begin{equation}
Y_{\mathrm{Sym}}\leq\frac{1}{100}\times\frac{\Omega_{\mathrm{DM}}h^2}{2.76\times 10^8}  \left(\frac{\GeV}{m_{\mathrm{DM}}}\right).
\label{sym}
\end{equation}
Varying the constraint on the symmetric component in a wide range around $1\%$ does not materially affect our conclusions concerning the required annihilation cross-section.

The annihilation cross-sections for the operators in Table \ref{tab1} connecting fermion DM and SM quarks to  $\OO(v^2)$ are \footnote{These expressions correct previous errors in the literature and have been checked both by hand and with {\tt{Feyncalc}} \cite{Feyncalc}.} 
%
\begin{align*}
\sigma_{\mathrm{An}}^{ \OO^\psi_{s}}  v  & =
 \frac{3v^2 m_{\mathrm{DM}}^2}{8\pi\Lambda^4}  \sum_q \left(1-\frac{m_q^2}{m_{\mathrm{DM}}^2}\right)^{3/2}, 
 \\[10pt]
\sigma_{\mathrm{An}}^{ \OO^\psi_{v}}  v  & = 
\frac{3m_{\mathrm{DM}}^2}{2\pi\Lambda^4} \sum_q  \left(1-\frac{m_q^2}{m_{\mathrm{DM}}^2}\right)^{1/2}
\Bigg[    \left(2+\frac{m_q^2}{m_{\mathrm{DM}}^2}\right)   + v^2 \left(\frac{8m_{\mathrm{DM}}^4-4m_q^2m_{\mathrm{DM}}^2+5m_q^4}{24m_{\mathrm{DM}}^2(m_{\mathrm{DM}}^2-m_q^2)}\right)  \Bigg],  
 \\[10pt]
  \sigma_{\mathrm{An}}^{ \OO^\psi_{a}} v  & =
 \frac{3m_{\mathrm{DM}}^2}{2\pi\Lambda^4}  \sum_q \left(1-\frac{m_q^2}{m_{\mathrm{DM}}^2}\right)^{1/2}
 \Bigg[ \frac{m_q^2}{m_{\mathrm{DM}}^2}  + v^2
 \left(\frac{8m_{\mathrm{DM}}^4-22m_q^2m_{\mathrm{DM}}^2+17m_q^4}{24m_{\mathrm{DM}}^2(m_{\mathrm{DM}}^2-m_q^2)}\right)\Bigg],
\\[10pt]
 \sigma_{\mathrm{An}}^{ \OO^\psi_{t}}  v & = 
  \frac{6m_{\mathrm{DM}}^2}{\pi\Lambda^4}  \sum_q \left(1-\frac{m_q^2}{m_{\mathrm{DM}}^2}\right)^{1/2} 
   \Bigg[ \left(1+2 \frac{m_q^2}{m_{\mathrm{DM}}^2} \right)
    +  v^2\left( \frac{4m_{\mathrm{DM}}^4-11m_{\mathrm{DM}}^2m_q^2+16m_q^2)}{24m_{\mathrm{DM}}^2(m_{\mathrm{DM}}^2-m_q^2)}\right)\Bigg],
    \\[10pt]
 \end{align*}
 \begin{align*}
\sigma_{\mathrm{An}}^{ \OO^\psi_{p}}  v & = 
\frac{3m_{\mathrm{DM}}^2}{2\pi\Lambda^4}  \sum_q \left(1-\frac{m_q^2}{m_{\mathrm{DM}}^2}\right)^{1/2}   
 \Bigg[ 1+\frac{v^2}{8} \left(  \frac{2m_X^2-m_q^2}{m_X^2-m_q^2}\right) \Bigg],
\\[10pt]
 \sigma_{\mathrm{An}}^{ \OO^\psi_{av}} v  & =
\frac{v^2 m_{\mathrm{DM}}^2}{4\pi\Lambda^4} \sum_q  \left(1-\frac{m_q^2}{m_{\mathrm{DM}}^2}\right)^{1/2} \left(2+\frac{m_q^2}{m_{\mathrm{DM}}^2} \right),    
\\[10pt]
\sigma_{\mathrm{An}}^{ \OO^\psi_{pt}} v  & =
\frac{6m_{\mathrm{DM}}^2}{\pi\Lambda^4} \sum_q  \left(1-\frac{m_q^2}{m_{\mathrm{DM}}^2}\right)^{1/2}
\Bigg[\left(1-\frac{m_q^2}{m_{\mathrm{DM}}^2}\right)
+ \frac{v^2}{24} \left(4+11 \frac{m_q^2}{m_{\mathrm{DM}}^2} \right)  \Bigg],  
 \\[10pt]
  \sigma_{\mathrm{An}}^{ \OO^\psi_{va}}  v & = 
 \frac{3m_{\mathrm{DM}}^2}{\pi\Lambda^4} \sum_q  \left(1-\frac{m_q^2}{m_{\mathrm{DM}}^2}\right)^{1/2}
 \Bigg[\left(1-\frac{m_q^2}{m_{\mathrm{DM}}^2}\right)
+ \frac{v^2}{24} \left(4+5 \frac{m_q^2}{m_{\mathrm{DM}}^2} \right)  \Bigg],  
\\
\end{align*}
while the annihilation cross-sections for operators with scalar DM are
\begin{align*}
 \sigma_{\mathrm{An}}^{ \OO^\phi_{s}}  v  &=
 \frac{3}{4\pi\Lambda^2} \sum_q \ \left(1-\frac{m_q^2}{m_{\mathrm{DM}}^2}\right)^{1/2}
  \Bigg[   \left(1-\frac{m_q^2}{m_{\mathrm{DM}}^2}\right)
 + \frac{3 v^2}{8} \frac{m_q^2}{m_{\mathrm{DM}}^2} \Bigg],
 \\[10pt]
 \sigma_{\mathrm{An}}^{ \OO^\phi_{v}}  v & =
 \frac{v^2m_{\mathrm{DM}}^2}{4\pi\Lambda^4} \sum_q\left(1-\frac{m_q^2}{m_{\mathrm{DM}}^2}\right)^{1/2} \left(2+\frac{m_q^2}{m_{\mathrm{DM}}^2}\right), 
 \\[10pt]
 \sigma_{\mathrm{An}}^{ \OO^\phi_{va}} v  & =  
 \frac{3 m_{\mathrm{DM}}^2}{\pi\Lambda^4} \sum_q \left(1-\frac{m_q^2}{m_{\mathrm{DM}}^2}\right)^{1/2} \Bigg[   \frac{m_q^2}{m_{\mathrm{DM}}^2}
 +
 \frac{v^2(7m_q^4-8m_q^2m_{\mathrm{DM}}^2+4m_{\mathrm{DM}})}{24m_{\mathrm{DM}}^2(m_{\mathrm{DM}}^2 -m_q^2)} \Bigg].
\end{align*}
%
In the above, the cross-sections corresponding to scalar or pseudoscalar  interactions ($ \OO^\psi_{s}$, $ \OO^\psi_{p}$ and $ \OO^\phi_{s}$) have been provided for the case of universal coupling, the cross-sections for the $m_q$-dependent couplings may be obtained by multiplying the cross-section (under the summation)  by a factor of $m_q^2/\Lambda^2$. To illustrate that the omitted CP violating operators are comparable to the CP conserving interactions we have included the operator $\OO^\psi_{pt}$.

These expressions are used to derive the maximum value of the scale $\Lambda$ permitted to ensure a suitable depletion of the symmetric component for each operator as a function of $m_{\mathrm{DM}}$, where we assume that only one operator is turned on at a time. Imposing the constraint on the symmetric abundance, eqn. (\ref{sym}), the results obtained are presented graphically in Figures \ref{Fig1}-\ref{Fig3}. The black curves present upper bounds for $\Lambda$ for each operator as a function of $m_{\mathrm{DM}}$. The resulting relic density curves may then be constrained through direct detection and collider limits, which we derive in the next section.

\section{Direct search constraints on contact operators}
\label{Sec3}

 The constraints on each contact operator in Table \ref{tab1} depend strongly on whether the corresponding scattering cross-section is spin-independent or spin-dependent. If the scattering cross-section is suppressed by powers of $v$ or $q$ then direct detection limits are severely weakened and, in fact, in all cases do not present any significant constraint on the allowed DM mass range. Rather, for the suppressed operators the leading bounds come from collider monojet searches, which we shall discuss shortly.

 The scattering cross-section per nucleon for each of the non-suppressed spin-independent interactions under consideration which couple universally to SM quarks\footnote{SI cross-section of operators which couple proportional to $m_q$ may be obtained by suitable rescalings; where we have assumed that $f_n\approx f_p$.} are \cite{Kurylov:2003ra, Fitzpatrick:2010em}
\begin{equation}
\begin{aligned}
\sigma_{\mathrm{SI}}^{ \OO^\phi_{s}} &\approx \frac{1}{4\pi\La^2}\frac{\mu_p^2}{m_{\mathrm{DM}}^2} f_p^2,
&\hspace{2cm}
\sigma_{\mathrm{SI}}^{ \OO^\phi_{v}} &\approx \frac{9}{4\pi\La^4}\mu_p^2, \\[5pt]
\sigma_{\mathrm{SI}}^{ \OO^\psi_{s}} &\approx \frac{1}{\pi\La^4}\mu_p^2 f_p^2, 
&\hspace{2cm}
\sigma_{\mathrm{SI}}^{ \OO^\psi_{v}} &\approx \frac{9}{\pi\La^4}\mu_p^2.
\end{aligned}
\end{equation}
where \(\mu_p=\frac{m_{\mathrm{DM}}m_p}{(m_{\mathrm{DM}}+m_p)}\) is the nucleon-DM reduced mass and \(f_{p}\) is the DM effective couplings to protons. For consistency with later calculations, we use the default values for the factor \(f_{p}\) from {\tt micrOMEGAs} \cite{Micromegas}. The non-suppressed scattering cross-sections for the spin-dependent operators of interest with universal couplings are given by
\begin{align}
 \sigma_{\mathrm{SD}}^{ \OO^\psi_{t}} \approx  4\times \sigma_{\mathrm{SD}}^{ \OO^\psi_{a}}  \approx \frac{16}{\pi\Lambda^4}\mu_p^2\left(\sum_q    \Delta^p_q \right)^2.
\end{align}
 The $\Delta^{p}_q$ account for the spin content of the nucleon and we use the values derived in \cite{Ellis:2008hf} from the COMPASS  \cite{Alekseev:2007vi} results:
$\left(\sum_q    \Delta^p_q \right)^2\approx0.32$.
Note that in the non-relativistic limit the form of the tensor operators $\bar{q}\sigma^{\mu\nu}q$ coincides with the axial-vector operator $\bar{q}\gamma^\mu\gamma^5 q$ up to a numerical factor \cite{Agrawal:2010fh}. From the scattering cross-sections given above we can use the current direct detection limits to place a lower bound on the scale $\Lambda$. 

\begin{figure}[t!]
\includegraphics[height=45mm,width=75mm]{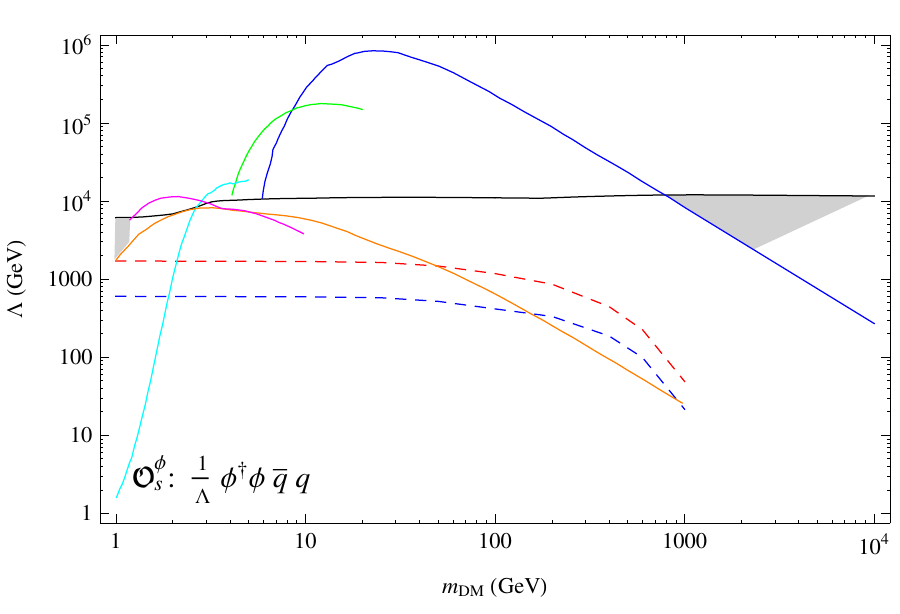}
\includegraphics[height=45mm,width=75mm]{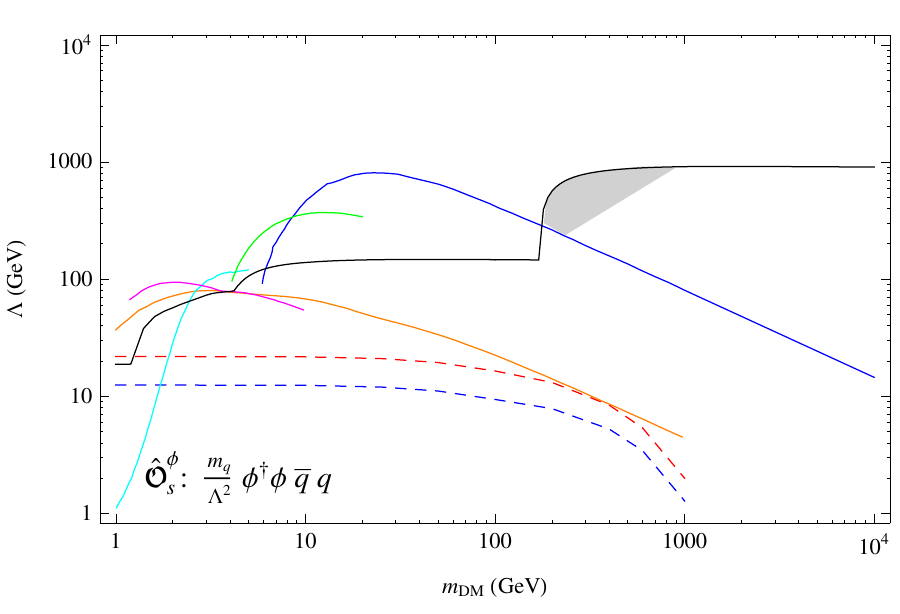}\\
\includegraphics[height=45mm,width=75mm]{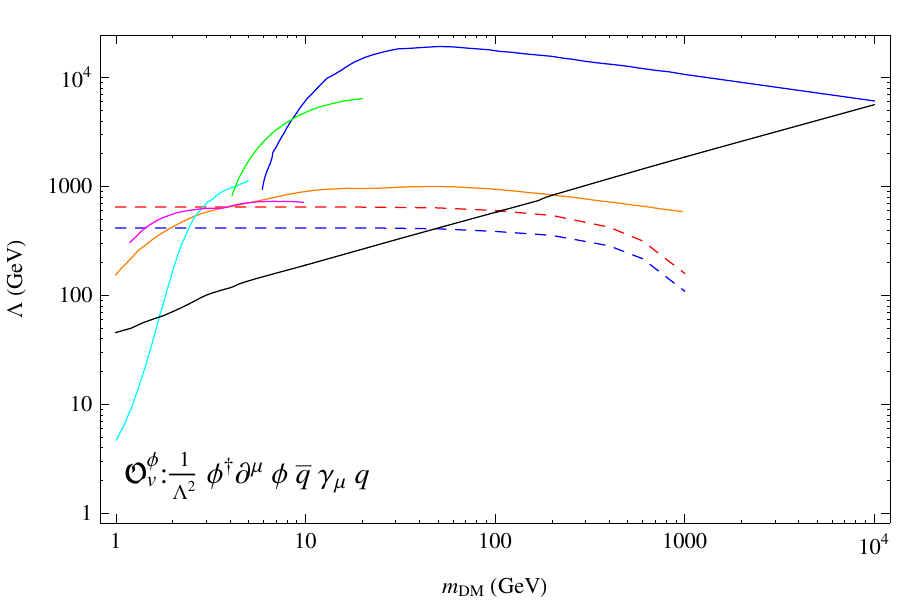}
\includegraphics[height=45mm,width=75mm]{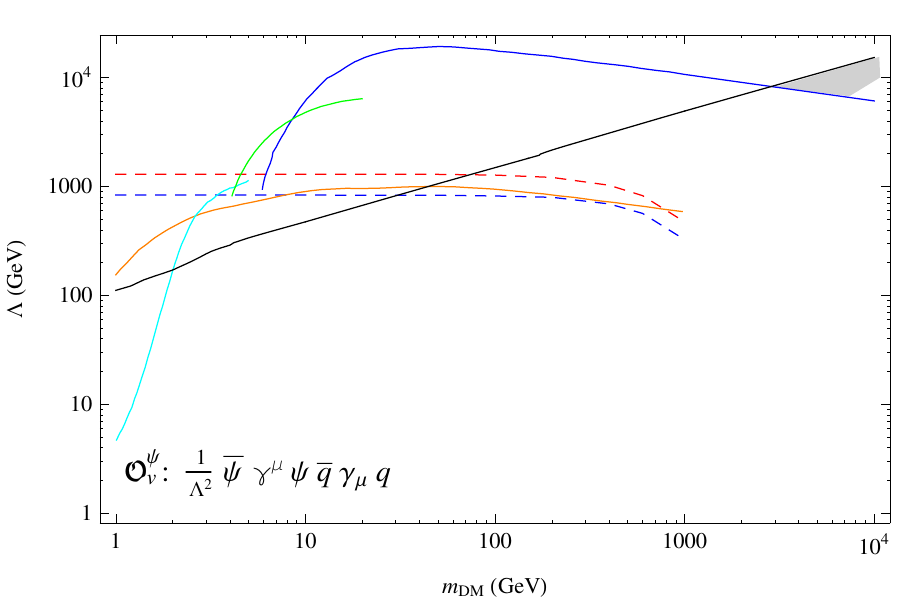}\\
\includegraphics[height=45mm,width=75mm]{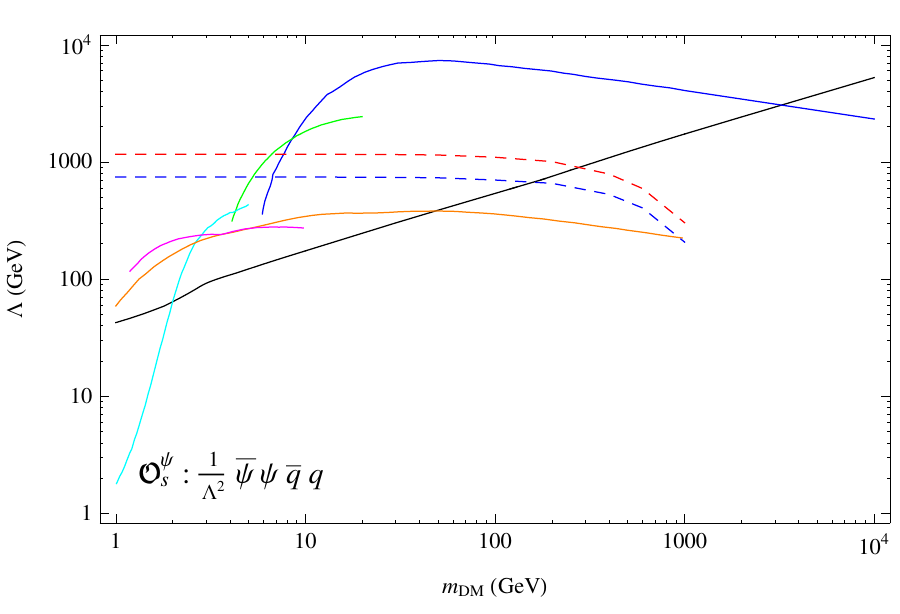}
\includegraphics[height=45mm,width=75mm]{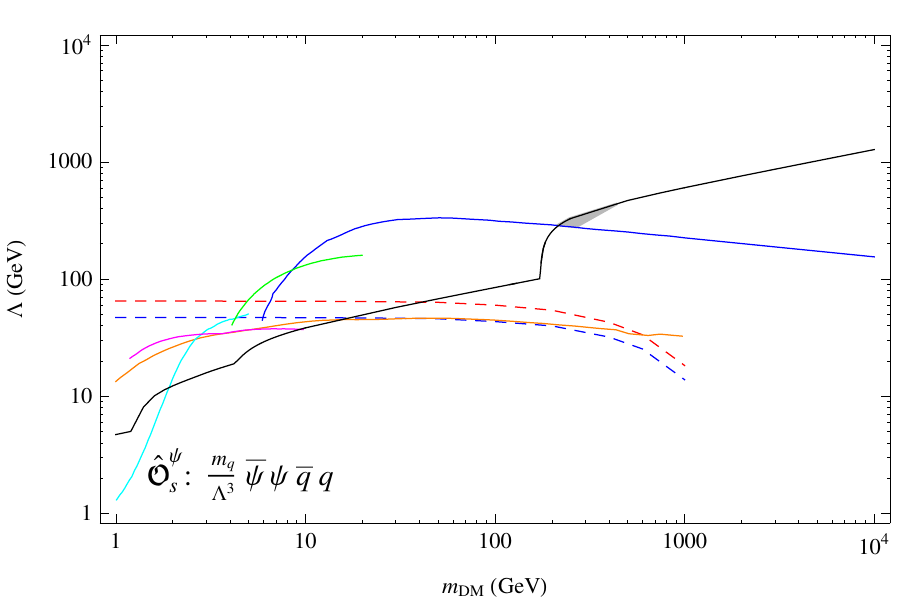}\\
\caption{Limits on the scale \(\Lambda\) for the effective operators from direct detection and monojets for scalar DM $\phi$ and fermion DM $\psi$, as a function of DM mass \(m_{\mathrm{DM}}\).   The black curve corresponds to the minimum annihilation cross-section necessary to reduce the symmetric component to \(1\%\). Constraints are from Xenon100 (blue), Xenon10 (green), CDMS  (cyan), CRESST (orange), DAMIC (magenta) and ATLAS  $1\,\fb^{-1}$ (red, dashed) and CMS $4.67\,\fb^{-1}$ (blue, dashed) monojet searches. For $m_{\mathrm{DM}}\gtrsim\Lambda $ effective operators no longer provide a good description and can not be reliably used to calculate the relic density requirements. Moreover, the monojet limits are no longer reliable much below $\Lambda\lesssim 100$ GeV due to the experimental cuts employed by ATLAS and CMS. Viable models of ADM employing the listed effective operators must lie in the shaded parameter regions. Note that contact operators due to scalar mediators are studied for both universal couplings to quarks and $m_q$-dependent couplings. We see that, except for the operator $\frac{1}{\Lambda}\phi^\dagger\phi\bar{q}q$ around $m_{\mathrm{DM}}\approx1$ GeV, successful models of ADM involving contact operators are excluded for $1~{\rm GeV}\lesssim m_{\mathrm{DM}} \lesssim 100~{\rm GeV}$, which includes the range in which ADM is most well motivated.} 
\label{Fig1}
\end{figure}

\begin{figure}[h!]
\includegraphics[height=43mm,width=75mm]{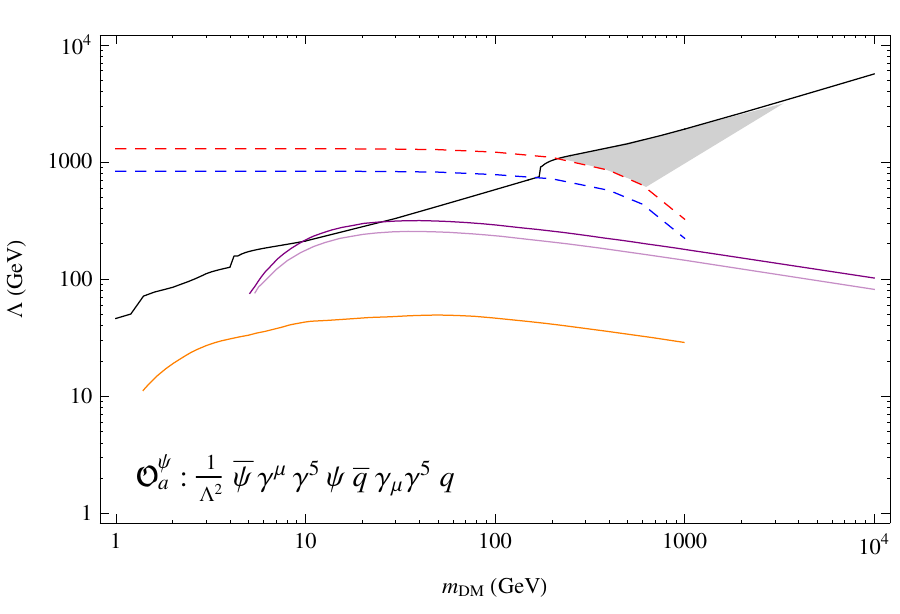}
\includegraphics[height=43mm,width=75mm]{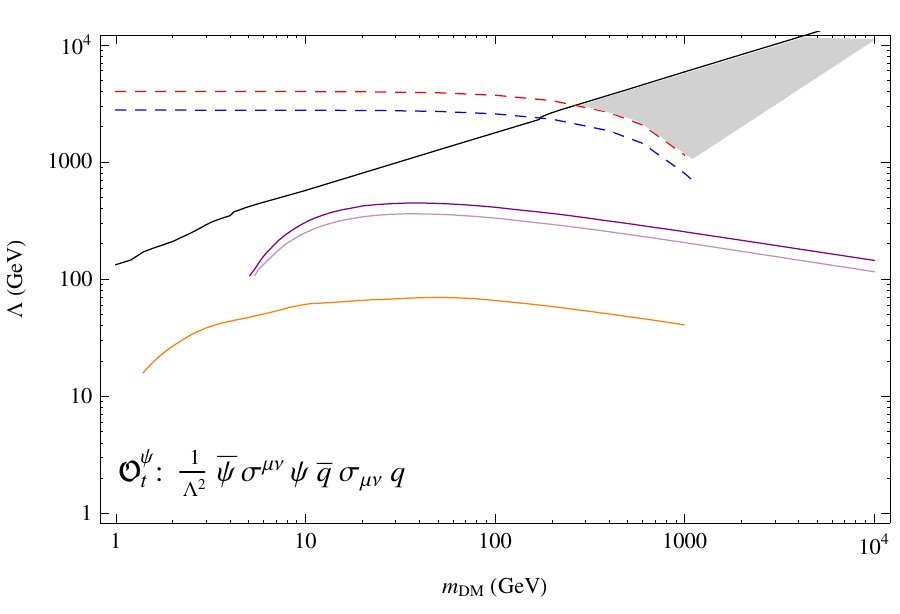}
\vspace{-5mm}
\caption{Limits on \(\Lambda\) for operators with spin-dependent direct detection cross-sections, viable parameter regions are shaded.  Constraints are from Simple (Stage 2: light purple; Combined: dark purple), CRESST (orange), ATLAS  $1\,\fb^{-1}$ (red, dashed) and CMS $4.67\,\fb^{-1}$ (blue, dashed).}
\label{Fig2}
\end{figure}

\begin{figure}[h!]
\vspace{-3mm}
\includegraphics[height=43mm,width=75mm]{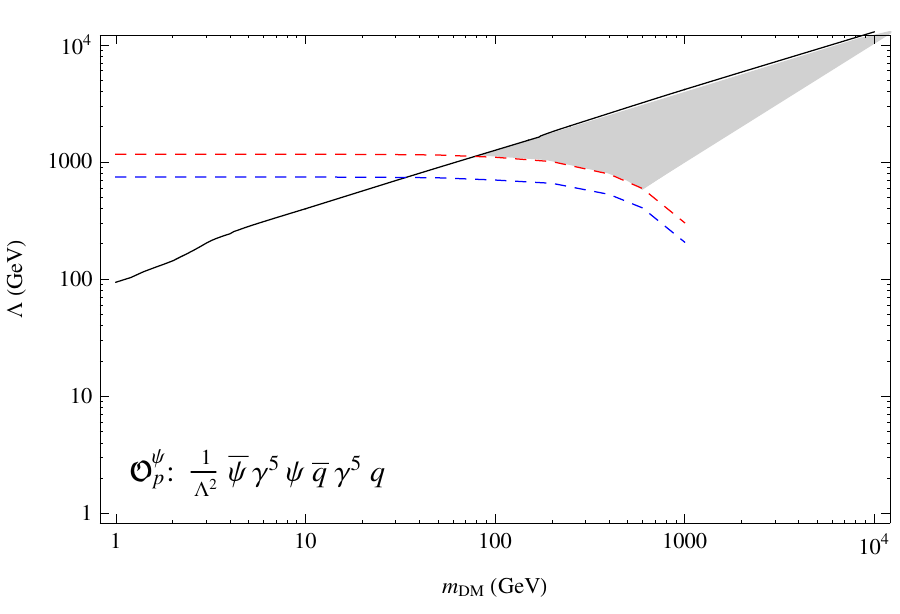}
\includegraphics[height=43mm,width=75mm]{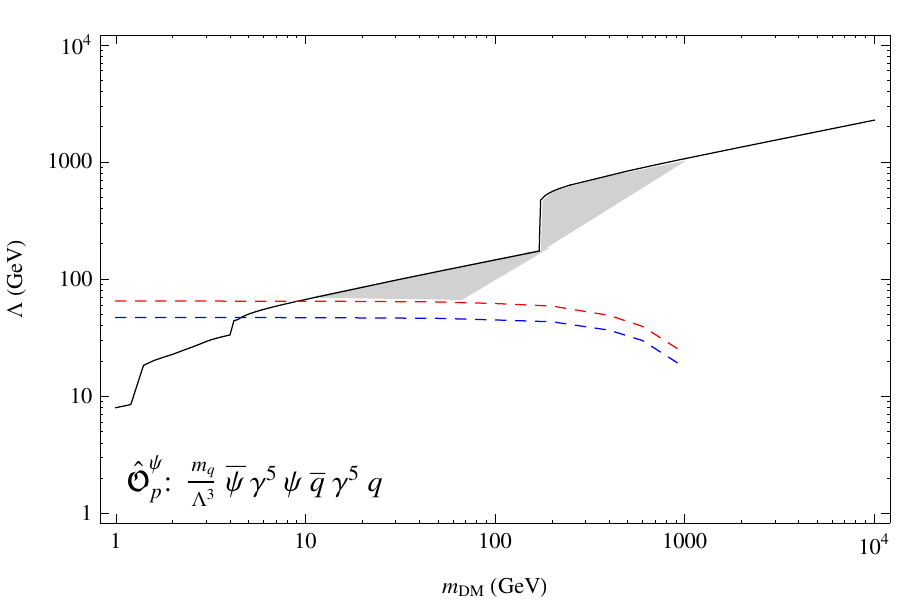}\\
\includegraphics[height=43mm,width=75mm]{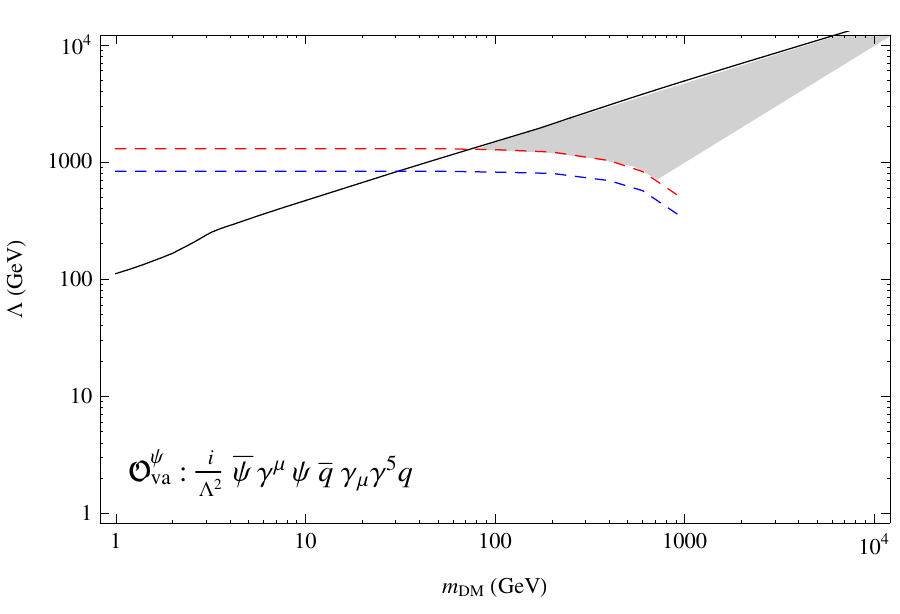}
\includegraphics[height=43mm,width=75mm]{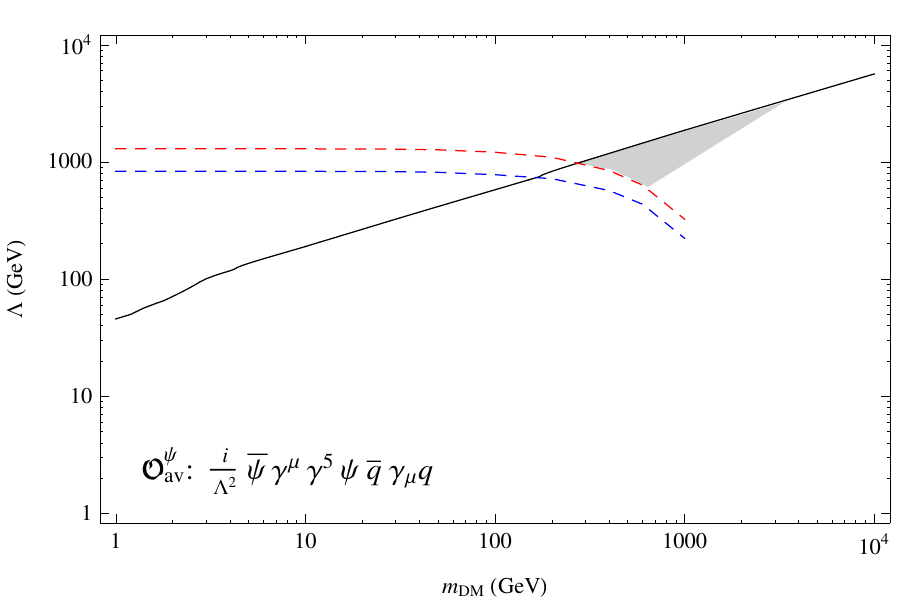}\\
\vspace{-3mm}
\includegraphics[height=43mm,width=75mm]{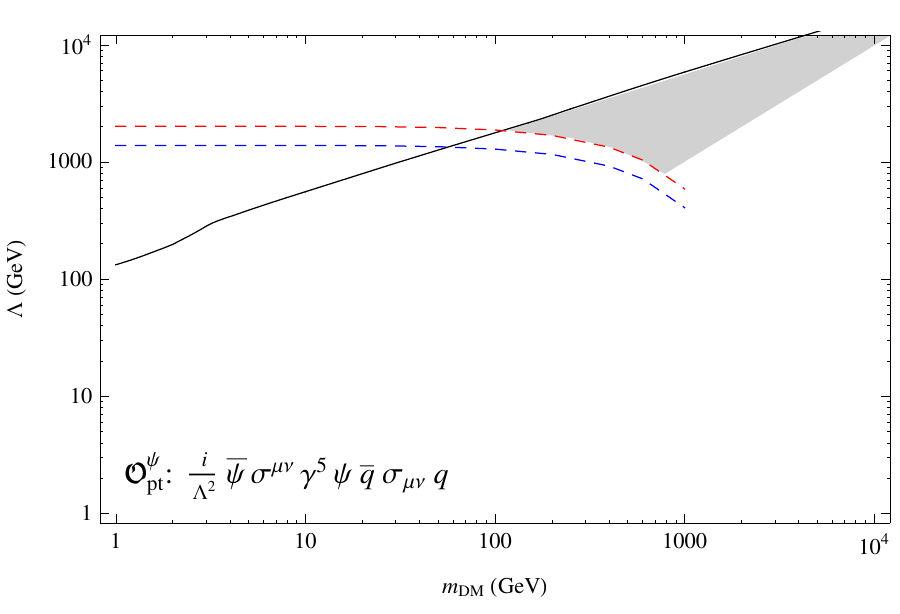}
\includegraphics[height=43mm,width=75mm]{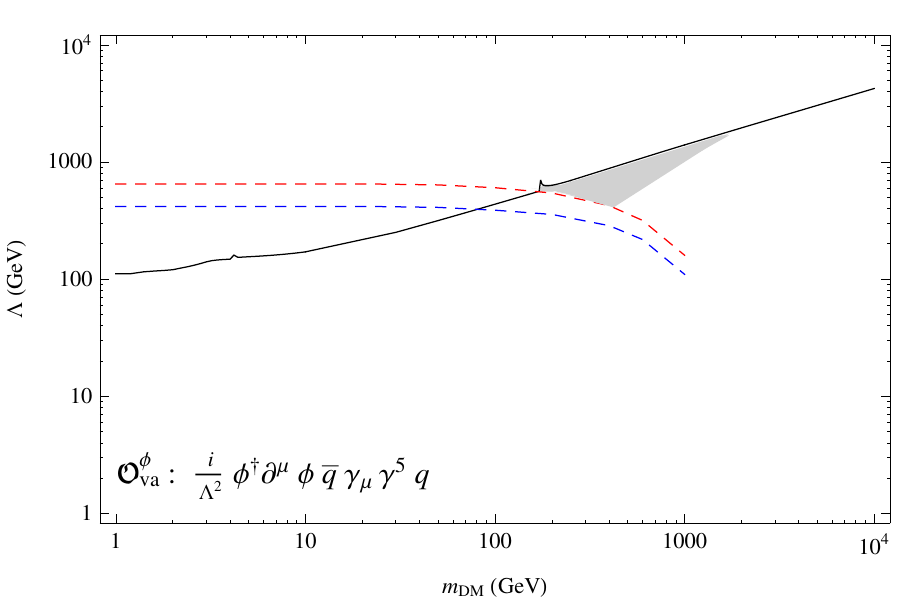}
\vspace{-2mm}
\caption{Limits on \(\Lambda\) for operators with $v$ or $q$ suppressed direct detection cross-sections, viable parameter regions are shaded. Limits are from ATLAS $1\,\fb^{-1}$ (red, dashed) and CMS $4.67\,\fb^{-1}$ (blue, dashed). The interesting ADM range $m_{\mathrm{DM}}\lesssim10$ GeV is excluded in all cases and, with the exception of the $\frac{m_q}{\Lambda^3}\bar{\psi}\gamma^5\psi\bar{q}\gamma^5q$ operator, this exclusion extends up to $m_{\mathrm{DM}}\lesssim100$ GeV.
}
\label{Fig3}
\end{figure}

The lower bounds placed on $\Lambda$ by direct detection are displayed in  Figures \ref{Fig1}-\ref{Fig3} as solid coloured curves. The leading limits on spin-independent interactions from direct detection experiments for $m_{\mathrm{DM}}\lesssim 1$ GeV come from CRESST \cite{Altmann:2001ax} (orange), while in the intermediate mass range  1 GeV$\leq m_{\mathrm{DM}}\leq$10 GeV the dominant constraints come from a combination of Xenon10 \cite{Angle:2011th} (green) CDMS \cite{Akerib:2010pv} (cyan) and DAMIC \cite{Barreto:2011zu} (magenta). For $m_{\mathrm{DM}}>10$ GeV the leading spin-independent limits are provided by the Xenon100 experiment \cite{Aprile:2011hi} (blue).  The strongest direct detection constraints on spin-dependent interactions are from Simple \cite{Felizardo:2011uw} (Stage 2: light purple; Combined: dark purple) for $m_{\mathrm{DM}}>5$ GeV  and from CRESST \cite{Altmann:2001ax} for lighter DM.




A complementary set of constraints on interactions between SM quarks and DM comes from collider experiments, in particular the recent LHC searches for jets and missing energy \cite{ATLAS,CMS}.\footnote{Note that LHC monophoton searches provide slightly weaker bounds to those from monojets \cite{CMS}.}  Operating at $\sqrt{s}=7$ TeV, the ATLAS and CMS collaborations have reported no apparent excess in the production of jets and missing energy with an integrated luminosity of $1\fb^{-1}$ and  $4.67\fb^{-1}$, respectively. The high $p_T$ ATLAS analysis \cite{ATLAS} selected events with a primary jet with $p_T>250$ GeV and pseudorapidity $|\eta|<2$, and $\slashed{E}_T>220$ GeV, events with a secondary jet with $p_T<60 $ GeV and $|\eta|<4.5$ were also permitted. The collaboration reported 965 events. The SM prediction for this process is $1010\pm75$ events. 
The ATLAS null result excludes any new contributions to the production cross-section greater than $0.01$ pb. 
The ATLAS monojet searches have been previously used to place limits on certain models of standard symmetric DM \cite{Rajaraman:2011wf,Fox:2011pm, Friedland:2011za, Shoemaker:2011vi, Razor}. Here we use the LHC monojet bounds to derive constraints on contact interactions between SM and DM states for the ADM scenario and employ the most recent results. The recent CMS search \cite{CMS} excludes new contributions to the production cross-section greater than $0.02$ pb and presents a comparable set of constraints to the ATLAS 1 $\fb^{-1}$ limits. The CMS analysis considered events with a primary jet with $p_T>110$ GeV and pseudorapidity $|\eta|<2.4$, and $\slashed{E}_T>350$ GeV, they also allowed events with a secondary jet with $p_T>30 $ GeV, provided the azimuth angle difference of the jets satisfied $\Delta\varphi<2.5$. CMS observed 1142 events in good agreement with the SM prediction of $1224\pm101$ events. 

To calculate the limits on contact operators coupling DM to quarks, we use the program {\tt CalcHEP} \cite{Pukhov:2004ca}. The SM Lagrangian is supplemented with each operator in separate instances and we study the total cross-section $pp\rightarrow j\bar{X}X$ where $X$ is the DM state and $j$ is a jet. To model the ATLAS search we apply the following cuts on the events $\slashed{E}_T>220$ GeV, $p_T>250$ GeV, and pseudorapidity $|\eta|<2$ and for CMS we take $\slashed{E}_T>350$ GeV, $p_T>110$ GeV, and pseudorapidity $|\eta|<2.4$. We do not consider events with additional jets in either cases. With these cuts we may reliably assume that the efficiency of the searches is close to 100\%. For each operator we determine the minimum value of $\Lambda$ for which the new contribution to the production cross-section is permitted by monojet searches. Since the contact operators are generally suppressed by multiple powers of $\Lambda$, $\OO(1)$ changes to the cross-section result in only small deviations in the limit on the scale $\Lambda$. The resulting limits are plotted in Figures \ref{Fig1}-\ref{Fig3} as dashed blue (CMS) and red (ATLAS) curves and supplement the constraints from direct detection. 

For $m_{\mathrm{DM}}\gtrsim\Lambda$ the contact operators no longer give a good description and one is required to consider the UV completion of the effective theory. Moreover, because of the cuts applied to the events by the experimental collaborations, the effective operators do not provide reliable bounds for monojet searches if the limit on $\Lambda$ is much below the $p_T$ cut $\sim100$ GeV.
 This is the case with the operator $\frac{m_q}{\Lambda^2}\phi\phi^{\dagger}\bar{q}q$, and the fermion DM, scalar and pseudoscalar $m_q$-dependent operator monojet bounds are close to the limit at which the effective theory breaks down. Consequently, these portal interactions should really be studied in the light mediator regime, which we discuss in detail in Section \ref{light}. Note, however, that the Tevatron monojet searches utilise much lower momentum cuts, requiring a primary jet with $p_T>10$ GeV, $|\eta|<1.1$, and $\slashed{E}_T>60$ GeV (90\% efficiency), and thus the contact operator description for monojet searches are reliable for $\Lambda\gtrsim 10$ GeV. Moreover, results from CDF \cite{Aaltonen:2012jb} lead to constraints which are roughly comparable to the CMS limits for $1~{\rm GeV}\lesssim m_{\mathrm{DM}} \lesssim 10~{\rm GeV}$. 

To correctly interpret Figures \ref{Fig1}-\ref{Fig3}, recall that the black curves provide an upper bound on $\Lambda$ to ensure suitably efficient annihilation of the symmetric component, whilst the coloured curves present lower bounds from searches. A given operator may present a viable ADM model only if  there exists some mass region for which a sufficiently low $\Lambda$ has not been excluded by direct search constraints. It is immediately apparent that there exists a tension between the experimental constraints and the upper bound on $\Lambda$ required for the efficient annihilation of the symmetric component over much of the model space.

ADM most naturally resides in the mass range $1~{\rm GeV}\lesssim m_{\mathrm{DM}} \lesssim 10~{\rm GeV}$. From the plots in Figures \ref{Fig1}-\ref{Fig3} we see that essentially no operators are viable in this preferred mass range. For scalar DM a small region of  parameter space remains at $m_{\mathrm{DM}}\approx 1$ GeV for the operator $\frac{1}{\Lambda}\phi\phi^{\dagger}\bar{q}q$. All of the other spin-independent non-suppressed operators are excluded by direct detection experiments up to at least $m_{\mathrm{DM}} =100 $ GeV. The strongest bounds on the remaining fermion DM operators in the ADM mass region are from collider searches, which are sufficient to exclude these operators over the mass range $1~{\rm GeV}\lesssim m_{\mathrm{DM}} \lesssim 10~{\rm GeV}$. With a small number of exceptions the common bound on the scale $\Lambda$ due to monojet searches is roughly $\Lambda\gtrsim 1$ TeV  for $m_{\mathrm{DM}}\lesssim100$ GeV and present stronger bounds than direct detection constraints in several cases. For the non-suppressed spin-independent operators,  since the ADM mass range is already  excluded by direct detection experiments, these collider limits do not impose significant new constraints.  In contradistinction the LHC searches provide virtually the sole limits on spin-dependent interactions and the velocity suppressed operators over the mass region of interest. 

On the other hand, inspecting Figures \ref{Fig1}-\ref{Fig3} we note that TeV scale ADM (in distinction to ADM in the natural mass range $1~{\rm GeV}\lesssim m_{\mathrm{DM}} \lesssim 10~{\rm GeV}$) allows most effective operators to remove the symmetric component efficiently without conflict with current detection limits. We have previously argued that if the asymmetries are equal, then DM should be roughly 5 GeV, however the TeV scale provides an alternative mass range for which the correct value of $\Omega_{\mathrm{DM}}$ can he realised \cite{Nussinov:1985xr,Barr:1990ca}. This is because for large masses DM production is Boltzmann suppressed by a factor $\exp\left(-\frac{m_{\mathrm{DM}}}{T}\right)$ for $m_{\mathrm{DM}}>T$. But, the TeV scale ADM scenario is rather less appealing, since the DM relic density is exponentially sensitive to changes in $m_{\mathrm{DM}}$. Alternatively, in sharing models the operator which transfers the asymmetry may be inefficient, and this can be used to generate large discrepancies between the baryon and DM asymmetries.  Whilst such constructions permit a wider range of $m_{\mathrm{DM}}$,  it as the mass of the DM rises it becomes increasingly difficult to explain the coincidence $\Omega_{\mathrm{DM}}/\Omega_B\approx 5$ via the DM asymmetry. Thus, although a wider range of ADM masses are certainly worth contemplating, since a natural expectation in many models of ADM is that the asymmetries in $X$ and $B$ should be comparable, we have focused much of our discussion on the case $m_{\mathrm{DM}}\sim m_p$.

A further source of limits on contact operators come from invisible quarkonium decays. Clearly, only bound states heavier than $2m_{\mathrm{DM}}$  can decay invisibly to the DM states. The principle constraints on the ADM region come from decays of the $\Upsilon(1S)$ mesons, as measured at the B-factories BaBar and CLEO \cite{BABAR}. 
However, we find that the limits on the contact operators of Table \ref{tab1}  coming from these invisible decay searches are not competitive with previous bounds from monojet searches or direct detection.  
There are also additional constraints on ADM from astrophysics \cite{Cumberbatch:2010hh,Zentner:2011wx,  Frandsen:2010yj, Iocco:2012wk, Davoudiasl:2011fj}. In particular for ADM models with scalar DM there are additional constraints from old compact stars \cite{Kouvaris:2010jy,McDermott:2011jp, Kouvaris:2011fi}.  These limits are sufficiently strong to exclude scalar ADM over a large mass range (including the natural ADM mass range), however this conclusion can be circumvented with further model building. For instance, these limits no longer apply if the scalar DM is not fundamental, but a composite state of fermions due to some new hidden sector strong dynamics, such that Fermi repulsion of the constituent states becomes important before the Chandrasekhar limit is reached \cite{Kouvaris:2011fi, Cline:2012is}.

In summary, the combination of direct detection and LHC limits excludes DM-quark contact interaction in the ADM mass region for all contact operators with minimal flavour structure. Moreover, with the exception of the pseudoscalar operator $\frac{m_q}{\Lambda^3}\bar{\psi}\gamma^5\psi\bar{q}\gamma^5q$, all of these contact operators are disallowed  by experimental searches for $1~{\rm GeV}\lesssim m_{\mathrm{DM}} \lesssim 100~{\rm GeV}$. The conclusion is striking: models of ADM with the DM at the natural mass scale (1-10 GeV) decaying directly to SM quarks via heavy mediators is completely excluded and only light mediators provide suitable portal interactions. The case of light mediators is not a trivial extension of the effective operator analysis since the results for the monojet and relic density calculations change drastically. We shall discuss the limits and model building opportunities of light mediators in the next section.


\section{Light mediator analysis}
\label{light}

A proper treatment of light mediators requires a careful analysis which includes effects of resonances and mass thresholds.  Mediators with mass comparable to, or less than, the momentum cuts employed in the collider searches can lead to a substantial weakening of the monojet bounds. Also, as noted previously, contact operators will not provide a faithful description of the monojet bounds if the limit on $\Lambda$ is much below the $p_T$ cut.
 In the presence of a light mediator $\eta$ the relic density depends upon the both the mass $m_{\mathrm{DM}}$ of the DM state $X$  and the mass of the mediator $m_\eta$.  The annihilation of the symmetric component is enhanced in the mass range $m_{\mathrm{DM}}\gtrsim m_\eta$, as the annihilation can proceed by the t-channel processes $\overline{X}X\rightarrow \eta\eta$. Moreover, when $m_\eta\approx2 m_{\mathrm{DM}}$ the annihilation via the s-channel process $\overline{X}X\rightarrow \eta\rightarrow$ SM is resonantly enhanced. Consequently, the case of GeV scale mediators must be carefully analysed and it is not sufficient to simply adjust the monojet limits as is often the approach taken in the previous literature.

In the remainder of this section we use the package {\tt micrOMEGAs} \cite{Micromegas} to study two well motivated examples and demonstrate the full effect of light mediators on relic density calculations and experimental bounds. To relate the features of the {\tt micrOMEGAs} results to the underlying physics, we study in the Appendix, a semi-analytic approach, following \cite{Griest:1990kh}.

\subsection{The light scalar-Higgs portal}
\label{higgs}

To illustrate this general formalism we shall consider a particular example of a scalar state which mediates interactions between fermion DM $\psi$ and SM quarks via mixing with the SM (or a SM-like) Higgs. In minimal models of DM it is expected that the mediator connecting the hidden and visible sectors should be a singlet under the SM gauge groups. In this case, mixing with the Higgs after electroweak symmetry breaking provides the dominant manner by which the mediator state couples to the SM quarks. After electroweak symmetry breaking the scalar mediator and the Higgs will mix giving physical mass eigenstates ($\eta$ and $h$) with interactions of $\eta$ with the SM quarks induced by this mixing \cite{Patt:2006fw, MarchRussell:2008yu}, which we parameterise
\bea
\mathcal{L}\supset\la_{X} \eta \bar{\psi} \psi +\sum_q (\theta\la^{\prime}  y_q) \eta \bar{q} q,
\eea
%
\begin{figure}[t!]
\begin{center}
\includegraphics[height=42mm,width=75mm]{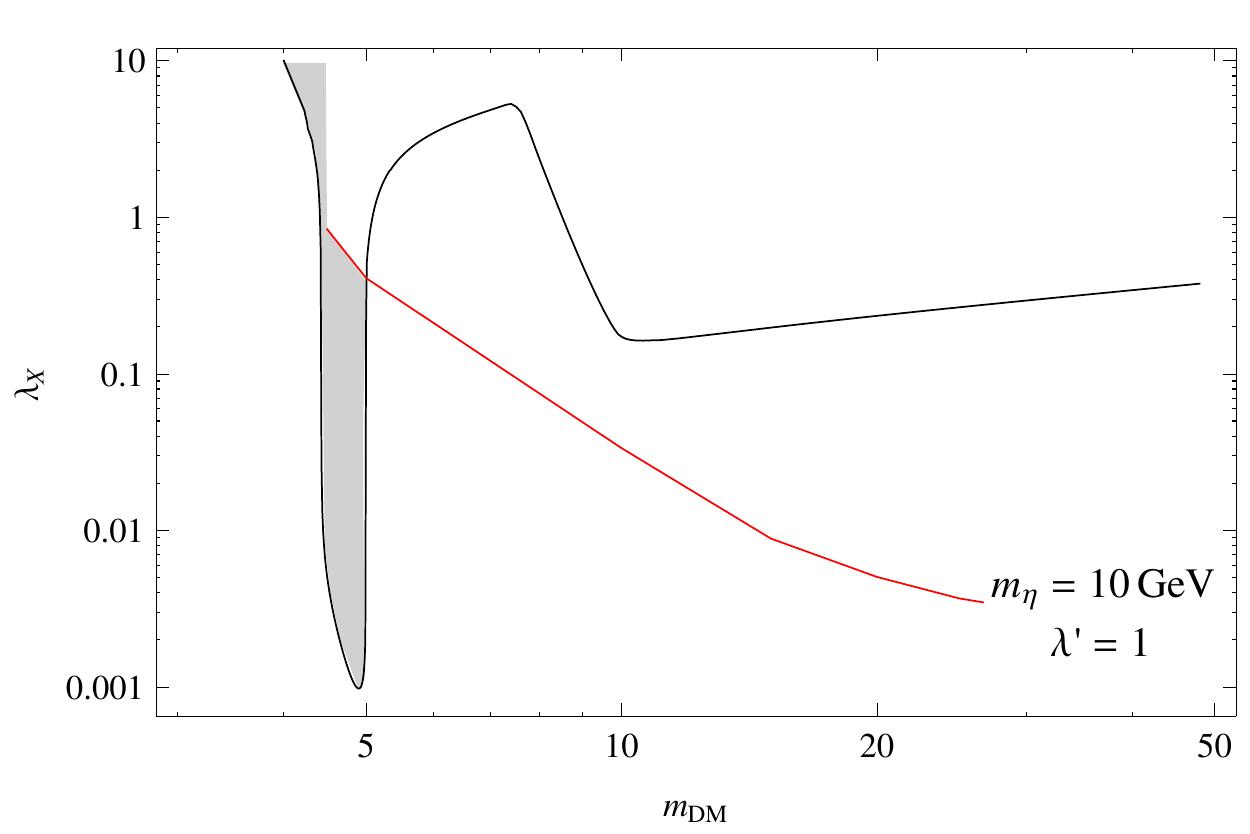}
\includegraphics[height=42mm,width=75mm]{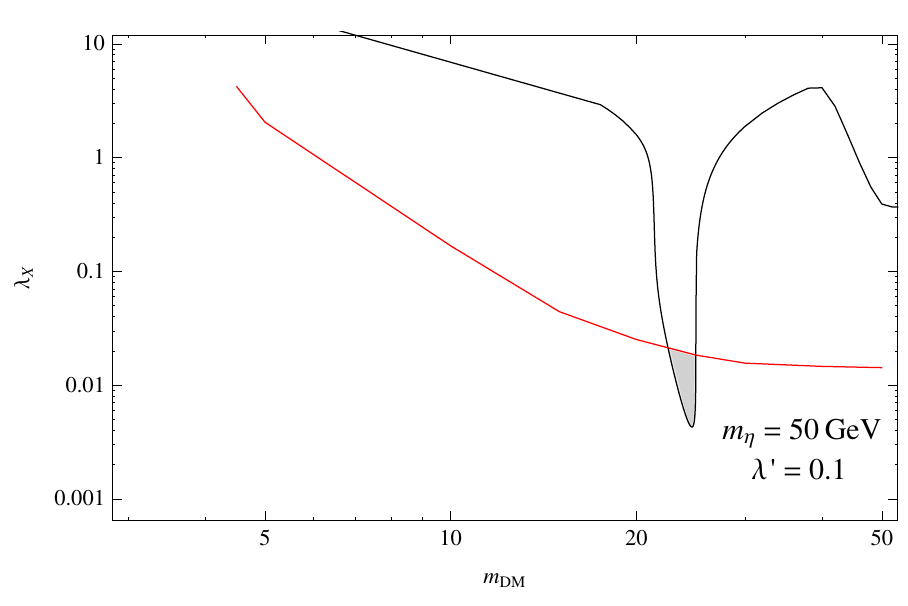}
\caption{Constraints on a light scalar mediator coupling  fermion DM to the visible sector via mixing with a SM-like Higgs with mass $m_h=125$ GeV.  The black curve shows the minimum DM-mediator coupling $\lambda_{X}$ required to efficiently annihilate the symmetric component of the DM. The red curve indicates the current combined spin independent direct detection bounds. Constraints from monojet searches are negligible.  The allowed parameter space is indicated by the shaded region. For mediators heavier than a few GeV only the resonance region survives the tension between direct detection limits and the requirement for efficient annihilation of the symmetric component.}
\label{Fig4}
\end{center}
\end{figure}
%
where $y_q$ are the standard model Yukawa couplings, $\theta\sim\frac{m_\eta}{m_h}$ is the mixing parameter and the product of these may be dressed by an additional coupling $\la^{\prime}$. We use {\tt micrOMEGAs} to calculate the required $\lambda_{X}$ necessary to obtain the asymmetry dominated relic density, as a function of DM mass $m_{\mathrm{DM}}$. See the Appendix for a semi-analytic treatment of ADM relic density in the presence of s-channel resonances $\bar{\psi}\psi\rightarrow\eta\rightarrow$ SM and t-channel threshold effects $\bar{\psi}\psi\rightarrow\eta\eta$. The results are shown in Fig. \ref{Fig4} for a scalar mediator with with parameters\footnote{LEP searches \cite{Barate:2003sz} for mixed Singlet-Higgs states constrain $\lambda^\prime\theta$ to be small and influence are choice of values for $\lambda^\prime$, however varying this parameter will only lead to rescaling of the results.} $\{m_\eta,~\lambda^\prime\}= \{10~\rm{GeV},~1\},\, \{50~\rm{GeV},~0.1\}$ and are accompanied by current direct detection constraints. For mediators heavier than a few GeV only the resonance region survives the tension between theoretical and experimental requirements. Moreover, because of the strength of the spin-independent direct searches even lighter mediators are constrained to the point where only a small part of the parameter space survives. For scalar mediators with mass $\lesssim$ few GeV, for $m_{\mathrm{DM}}\sim m_\eta$, DM annihilation into $\eta$ pairs (which later decay to SM states) is allowed by direct detection constraints and presents a possible mechanism. However, mediators with $m_\eta\lesssim$ 10 GeV or DM with $m_{\mathrm{DM}}\lesssim$ 5 GeV will lead to additional limits, most prominently searches for quarkonium decays to invisible states \cite{BABAR}. If the experimental hints of a SM-like Higgs boson at approximately 125 GeV are confirmed, then this portal interaction will be constrained by subsequent precision measurements of Higgs decay rates \cite{Djouadi:2011aa, Englert:2011yb, LopezHonorez:2012kv}.

\subsection{The light pseudoscalar portal}

\label{sec2.5}

\begin{figure}[t!]
\includegraphics[height=42mm,width=75mm]{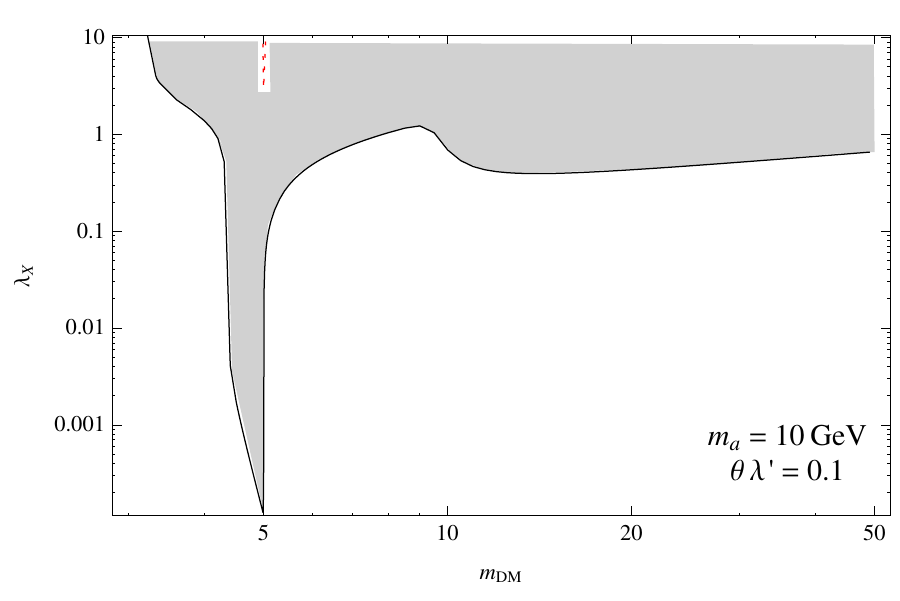}
\includegraphics[height=42mm,width=75mm]{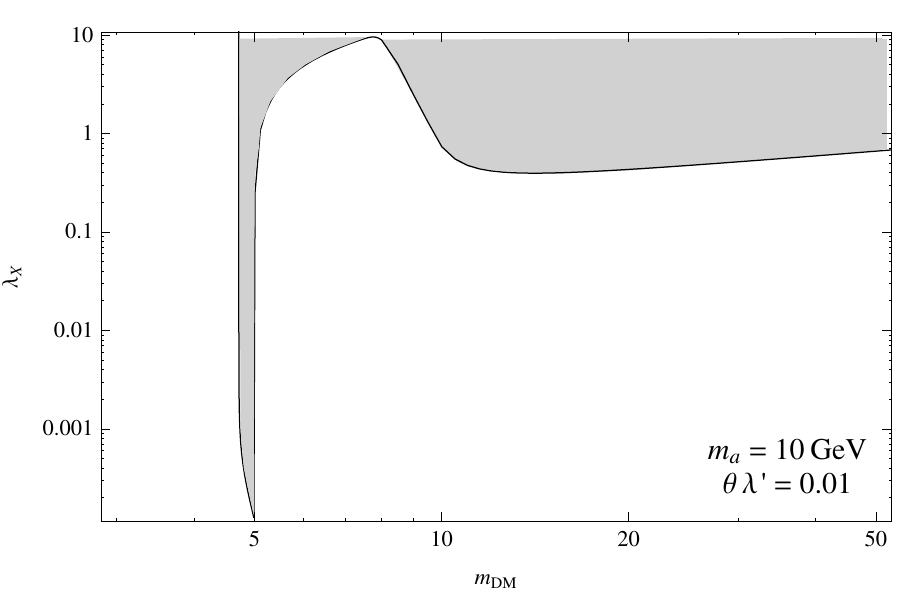}\\
\caption{
Constraints on a pseudoscalar mediator coupling fermion DM to the visible sector.  The solid curve shows the required hidden sector coupling in order to efficiently annihilate the symmetric component. The dashed curve visible near 5 GeV (the resonance region) indicates the LHC monojet limits, as can be seen these do not present significant constraints on the model, this is in contrast to the corresponding contact operator $\OO^\psi_{p}$ studied in Section \ref{Sec3}. Moreover, as the scattering cross-section is suppressed by $q^4$ there are essentially no limits from direct detection. The allowed parameter space is indicated by the shaded region.} 
\label{Fig5}
\end{figure}

It was seen in Section \ref{Sec3} that heavy mediators are greatly disfavoured in natural implementations of ADM. Pseudoscalars are an ideal candidate for light mediator states and present a particularly interesting case as the constraints from direct detection (suppressed by $q^4$) are negligible. We parametrise the portal interaction connecting fermion DM $\psi$  with SM quarks via a pseudoscalar mediator as follows
\begin{equation}
\mathcal{L} \supset i\lambda_{X} a \overline{\psi}\gamma^5\psi  + \sum_q i(\theta\lambda^{\prime} y_q) a \overline{q}\gamma^5fq,
\label{LL}
\end{equation}
where, as previously, $y_q$ are the standard model Yukawa couplings and $\theta$ is the mixing parameter. Such a scenario could occur via mixing with a CP-odd state of an extended Higgs sector (e.g. the $A^0$ of the MSSM).

In Fig. \ref{Fig5} we display (black solid curve) the minimum $\lambda_{X}$ required for efficient annihilation of the symmetric component, as a function of $m_{\mathrm{DM}}$, for a portal interaction due to a 10 GeV pseudoscalar mediator $a$ for $\theta\lambda^{\prime}=0.1,\,0.01$ as the black solid curve. By fixing the coupling of the mediator to SM quarks, the size of the DM coupling $\lambda_{X}$ to the mediator required for efficient annihilation of the symmetric component is simply a function of the DM mass and the mediator mass. This allows us to present the effects of varying the parameters in a clear manner, which can be readily compared with the previous example of the light scalar-Higgs portal. The leading constraints on the light pseudoscalar portal come from the LHC monojet searches. We use {\tt CalcHEP} to determine the LHC monojet bounds following the same procedure as in Section \ref{Sec3} and these limits are presented as the red dashed curve in Fig. \ref{Fig5}.

We note that both the monojet limits and the DM relic density are very sensitive to the DM mass, with both curves featuring resonances and mass threshold effects. In contrast to the effective operator analysis, where ADM was excluded up to 100 GeV by collider searches, the monojets limits for light mediators $m_a\lesssim$100 GeV are weakened to the extent that for natural couplings they present no appreciable constraint. Indeed the monojet limits are only visible in the plots of Fig. \ref{Fig5} in the resonance region $m_a\approx2 m_{\mathrm{DM}}$.

For $m_a\neq2m_{\mathrm{DM}}$ the DM annihilation is via a virtual mediator and this leads to a large suppression in the cross-section, thus for natural values of the parameter the mediator and DM masses must typically be arranged such that the annihilation of the symmetric component proceeds via resonant s-channel processes or t-channel pair production of the mediator states. In particular, if the DM is heavier than the mediator then there will generally be unconstrained parameter space in which efficient annihilation of the symmetric component to the visible sector can be achieved. Notably, light pseudoscalars states are very well motivated, as they typically occur as pseudo-Nambu-Goldstone Bosons.


\section{Concluding remarks}
\label{Sec4}

If one begins from the well motivated assumptions that ADM with mass $m_{\mathrm{DM}}\sim m_p$ annihilates dominantly to SM quarks, and the flavour structure of such DM-quark couplings is minimal, then we have shown (Figures \ref{Fig1}-\ref{Fig3}) that the current experimental limits allow one to make decisive statements regarding the nature of the hidden sector. 
In Fig. \ref{Fig1}  it can be seen that contact operators with spin-independent non-suppressed direct detection cross-sections are excluded for $\Lambda\lesssim$1 TeV with universal couplings and up to 200 GeV in the case of $m_q$ dependent couplings. The one exception is the operator $\frac{1}{\Lambda}\phi^\dagger\phi\bar{q}q$ for $m_{\mathrm{DM}}\lesssim1$ GeV. We expect that this small window will be explored by ATLAS and CMS in their 2012 run, where $\gtrsim10\,\fb^{-1}$ will be collected by each experiment. Contact operators with spin-dependent non-suppressed direct detection cross-sections are displayed in Fig. \ref{Fig2} and it is seen that these are excluded by monojet searches up to $\Lambda\sim300$ GeV. Finally, the suppressed direct detection cross-sections displayed in Fig. \ref{Fig3} are excluded for $\Lambda\lesssim$100 GeV with the exception of  the pseudoscalar operator $\frac{m_q}{\Lambda^3}\bar{\psi}\gamma^5\psi\bar{q}\gamma^5q$ which is ruled out in the range $\Lambda\lesssim$10 GeV. 

Collating these results we conclude that most contact operators are very strongly constrained and, moreover, that all of the operators are essentially excluded in the interesting, motivated ADM mass region $1~{\rm GeV}\lesssim m_{\mathrm{DM}} \lesssim 10~{\rm GeV}$.
However, the effective operator description is no longer valid for mediators lighter than $\lesssim$100 GeV, at which point  resonance and threshold effects become important in determining the exclusion limits. As can be seen in Figures \ref{Fig4} and \ref{Fig5}, if the mediator states are light then successful models of ADM can be constructed.  Hence, because of the strong constraints on contact operators, it is required that the mediator connecting the visible and hidden sectors be relatively light in models of ADM conforming with our natural assumptions in order to efficiently annihilate the symmetric component whilst remaining consistent with search constraints. Moreover, it can be seen in Figures \ref{Fig4} and \ref{Fig5} that for mediators with mass $\sim10-100$ GeV there remain strong limits in many cases and to satisfy these constraints it is often required that the DM and mediator masses are arranged such that the DM annihilation is resonantly enhanced.

Some caveats to the conclusion presented above are in order. Most prominently, the DM may couple dominantly to leptons. Although such a coupling would go against our expectations from models of flavour and from grand unified theories, if the DM interacts more strongly with leptons compared to quarks, this would greatly relax the monojet bounds and the constraints from DM direct detection \cite{Fox:2008kb}. The dominant bounds would now come from monophoton experiments and DM experiments which do not veto on electron recoil, such as DAMA/LIBRA \cite{Bernabei:2010mq}.  Alternatively one may attribute a non-minimal flavour structure to the interaction with quarks, for instance, the DM could couple in an isospin violating fashion \cite{Feng:2011vu}. There exist studies of the collider limits in both the leptophilic \cite{Fox:2011fx}  and isospin violating \cite{Rajaraman:2011wf} scenarios with reference to conventional models of DM. Whilst possible in the context of ADM models these variant theories require further model building and a dedicated study.

The introduction of new light states is perhaps the most interesting and well motivated manner to avoid the strong constraints presented here. If the DM belongs to an extended hidden sector, containing lighter states into which the symmetric component can annihilate, then the limits from both direct search constraints and cosmological requirements are greatly ameliorated. Indeed, entire hidden sectors are well motivated in many models of physics beyond the standard model. In particular, it has been argued that the topological complexity of generic string compactifications leads to the expectation of many hidden sectors sequestered from the SM states \cite{Harling:2008px,  Chen:2006ni, Hebecker:2006bn, Berg:2010ha}. Studies of hidden sector cosmology and collider phenomenology exist in the literature \cite{Cohen:2010kn,  Morrissey:2009ur, Cheung:2010gj, Cheung:2010gk} and this is also related to the hidden valley scenario \cite{Strassler:2006im}.

The findings of this paper show that experimental searches strongly motivate the existence of additional hidden light degrees of freedom in models of ADM, either to mediate interactions between hidden and visible sector or into which the DM can annihilate. If the mediator state is made sufficiently light then portal interactions can provide mechanisms for ADM that both provide efficient annihilation of the symmetric component and satisfy limits from monojet and direct detection searches. However, depending on the mass and nature of the mediator there are further model dependent constraints on the couplings of these states, such as invisible quarkonium decay, beam dump experiments, invisible Z decay width, electroweak precision measurements, muon $g-2$, atomic parity violation, and also astrophysical processes. The cosmology, phenomenology and model building opportunities presented by extended hidden sectors of ADM is the focus of our forthcoming companion paper \cite{companion}.


\section*{Acknowledgements}

JU would like to thank Felix Kahlhoefer for useful discussions and Artur Apresyan for helpful correspondence. 
JMR is supported in part by both ERC Advanced Grant BSMOXFORD 228169, and acknowledges support from EU ITN grant UNILHC 237920 (Unification in the LHC era).
JU is grateful for partial support from the Esson Bequest, Mathematical Institute, Oxford.
SMW thanks the Oxford physics department for hospitality and the Higher Education Funding Council for England and the Science and Technology Facilities Council for financial support under the SEPNet Initiative. SMW is also grateful for financial support from the IPPP.


\appendix
\section{Semi-analytic analysis of light mediators}

We shall make some remarks regarding the source of the features exhibited in the relic density calculations with light mediators using a semi-analytic analysis.  The introduction of light mediators leads to resonance and mass threshold effects and the dominant process which determines the relic density becomes strongly dependent upon the DM mass. Note also that many models, in particular supersymmetric models, may lead to co-annihilation effects which must be accounted for but this is beyond the scope of this short appendix. In correctly specifying the annihilation integral $J$ for light mediators, we shall follow the work of Griest and Seckel \cite{Griest:1990kh}.

With a light mediator $\eta$ the standard expansion of $\langle \sigma  v \rangle$ used for contact operators, eqn. (\ref{sig}), is not valid over the whole mass range of the DM, $X$. There are a number of notable ranges of the DM mass in which the annihilation of the symmetric component is enhanced. In the DM mass range $m_{\mathrm{DM}}\gtrsim m_\eta$ the annihilation can proceed by the t-channel process $\overline{X}X\rightarrow \eta\eta$. The second region of interest is when $m_\eta\approx2 m_{\mathrm{DM}}$, in which case annihilation via s-channel $\overline{X}X\rightarrow \eta\rightarrow \overline{q}q$ is resonantly enhanced.

To model the effects of resonant production we factor out the pole factor $P(v^2)$ from the cross-section
\begin{equation}
P(v^2)=\left[\left(1-\frac{\left(2m_{\mathrm{DM}}/m_\eta\right)^2}{1-\frac{v^2}{4}}\right)^2+\left(\frac{\Gamma_\eta}{m_\eta}\right)^2\right]^{-1}.
\label{Pv}
\end{equation}
Whilst more sophisticated approximations are discussed in  \cite{Griest:1990kh}, this will suffice for our purposes.
Following this we can write the thermally averaged cross-section in the mass region $m_{\mathrm{DM}}\lesssim m_\eta$ as follows
\begin{equation}
\langle \sigma  v \rangle_{m_{\mathrm{DM}} \lesssim m_\eta}\simeq \left(a_s+\frac{6b_s}{x} \right)\times P(0),
\end{equation}
where $a_s$ and $b_s$ are the coefficients from the velocity expansion of the cross-section corresponding to s-channel  $\overline{X}X\rightarrow \eta\rightarrow \overline{q}q$.  It is clear from inspection of eqn (\ref{Pv}) that the cross-section will be peaked at $m_{\mathrm{DM}}\approx m_\eta/2$. In which case the annihilation integral, appearing in eqn. (\ref{YX})  may be expressed as follows
\begin{equation}
J_{m_{\mathrm{DM}}\lesssim m_\eta} \simeq \frac{\left(a_s+\frac{3b_s}{x_F} \right)\times P(0)}{x_F}.
\end{equation}

In the mass region $m_\eta\gtrsim m_{\mathrm{DM}}$ the states $\eta$ can be pair produced and present the dominant annihilation channel. For DM states with mass $m_{\mathrm{DM}}\approx m_\eta$ the final state velocity $v_f$ becomes important and can not be approximated as unity in the velocity expansion of the cross-section
\begin{equation}
\sigma v \approx (a_t+b_tv^2)v_f.
\end{equation}
The quantity $v_f$ can be expressed as follows
\begin{equation}
v_f=\sqrt{1-\left(\frac{m_\eta}{m_{\mathrm{DM}}}\right)^2\left(1-\frac{v^2}{4}\right)}.
\end{equation}
We may make a velocity expansion of $v_f$ 
\begin{equation}
v_f\approx\sqrt{1-\frac{m_\eta^2}{m_{\mathrm{DM}}^2}}\left(1+\frac{m_\eta^2v^2}{8(m_{\mathrm{DM}}^2-m_\eta^2)}\right).
\label{vf}
\end{equation}
Note that for $m_\eta\ll m_{\mathrm{DM}}$ the $v_f\approx1$ and we recover the standard velocity expansion for the cross-section. Furthermore, observe that the velocity expansion for $v_f$ breaks down for $m_{\mathrm{DM}}\approx m_\eta$. The annihilation integral for $m_{\mathrm{DM}}>m_\eta$ and away from the threshold region $m_{\mathrm{DM}}\approx m_\eta$, may be approximated thus
\begin{equation}
J_{m_{\mathrm{DM}}\gtrsim m_\eta } \simeq
\frac{\left(a_tI_a+\frac{3b_tI_b}{x_F} \right)}{x_F},
\end{equation}
where $a_t$ and $b_t$ are the coefficients from the velocity expansion of the cross-section corresponding to t-channel $\overline{X}X\rightarrow \eta\eta$.  The factors $I_{a,b}$ account for the final state velocities of the produced states and are given by:
\begin{align}
I_a =\frac{x_F}{a_t}\int_{x_F}^\infty\frac{\langle a_t v_f\rangle}{ x^{2}}\mathrm{d} x,
\hspace{1cm}
\mathrm{and}
\hspace{1cm}
I_b =\frac{2x_F^2}{b_t}\int_{x_F}^\infty\frac{\langle b_t v^2 v_f\rangle}{ 6x^{2}}\mathrm{d} x.
\label{I}
\end{align}
 The explicit form of the factors $I_{a,b}$ given in eqn. (\ref{I}) depend on $v_f$. Away from the threshold region  the approximation in eqn. (\ref{vf}) is valid, however near the threshold one must use the exact form for $v_f$. Moreover, since the interacting states have non-zero velocities, processes which would be forbidden at zero velocity may be accessible and lead to the dominant decay channels. We construct the total annihilation integral as a piecewise function:
\begin{equation}
J=
\left\{
 \begin{array}{lll}
J_{m_{\mathrm{DM}}\lesssim m_\eta } &\qquad m_{\mathrm{DM}}<0.97 \times m_\eta\\
J_{\mathrm{Threshold}} &\qquad 0.97 \times m_\eta \leq  m_{\mathrm{DM}} \leq \times1.03 m_\eta\\
J_{m_{\mathrm{DM}}\gtrsim m_\eta } &\qquad  1.03 \times m_\eta < m_{\mathrm{DM}}
 \end{array}
 \right. ,
\end{equation}
where $J_{\mathrm{Threshold}}$ is identical in form to $J_{m_{\mathrm{DM}}\gtrsim m_\eta } $ but the $I_{a,b}$ factors are evaluated using the exact form for $v_f$. The factors $I_{a,b}$ are given explicitly in \cite{Griest:1990kh} for all cases.

To calculate the relic density with light mediators we must use the corrected annihilation integral in eqn. (\ref{YXJ}):
\begin{equation*}
Y_{\mathrm{DM}} = \frac{\eta_{\mathrm{DM}}}{1-\exp\left[-\eta_{\mathrm{DM}} J \omega \right]},
\hspace{15mm}
Y_{\bar{\mathrm{DM}}} = \frac{\eta_{\mathrm{DM}}}{\exp\left[ \eta_{\mathrm{DM}} J \omega\right]-1}.
\end{equation*}
Then, as previously, we require that the asymmetric component alone  accounts for the relic density by the following requirement on the symmetric component:
\begin{align*}
Y_{\mathrm{Sym}}\leq \frac{1}{100} \times \frac{\Omega_{\mathrm{DM}} h^2}{2.76\times 10^8} \left(\frac{\GeV}{m_{\mathrm{DM}}}\right).
\end{align*}
At this stage, for a given model, one can numerically evaluate the analytic forms above. This semi-analytic computation gives comparable results to those obtained from {\tt micrOMEGAs}. Consequently, the features appearing in the light mediator analysis of Section \ref{light} can be well understood by comparison to the semi-analytic approach.


\end{document}